\documentclass[twocolumn,showpacs,preprintnumbers,amsmath,amssymb]{revtex4}

\usepackage{graphicx}
\usepackage{dcolumn}
\usepackage{bm}

\newcommand \bra[1]{\left< {#1} \,\right\vert}
\newcommand \ket[1]{\left\vert\, {#1} \, \right>}

\newcommand{\bea}{\begin{eqnarray}}
\newcommand{\eea}{\end{eqnarray}}
\newcommand{\simgt}{\hbox{ \raise3pt\hbox to 0pt{$>$}\raise-3pt\hbox{$\sim$} }}
\newcommand{\simlt}{\hbox{ \raise3pt\hbox to 0pt{$<$}\raise-3pt\hbox{$\sim$} }}

\newcommand{\LQ}{\Lambda_{\rm QCD}}

\begin{document}

\preprint{TU--856, KEK--TH-1339}

\title{Static QCD potential at three--loop order}

\author{C.~Anzai$^\dagger$, Y.~Kiyo$^\ast$ and Y.~Sumino$^\dagger$}
\affiliation{
$^\dagger$Department of Physics, Tohoku University,
Sendai, 980--8578 Japan
\\
\\
$^\ast$Theory Center, KEK, Tsukuba, Ibaraki 305-0801, Japan
}%

\date{\today}

\begin{abstract}
We compute the purely gluonic contribution to the static QCD potential
at three--loop order.
This completes the computation of the static potential at this order.
\end{abstract}

\maketitle

For more than 30 years, the static QCD potential $V_{\rm QCD}(r)$
has been studied extensively
for the purpose of elucidating the nature of the interaction between heavy
quark and antiquark.
Generally, $V_{\rm QCD}(r)$ at short-distances can be computed accurately
by perturbative QCD.
On the other hand,
the potential shape at long-distances should be determined by
non-perturbative methods, such as
lattice simulations or phenomenological potential-model analyses or
computations based on string-inspired models.

Computations of $V_{\rm QCD}(r)$ in perturbative QCD has a long
history.
At tree-level, $V_{\rm QCD}(r)$ is merely a Coulomb potential,
$-C_F\alpha_S/r$ ($C_F=4/3$ is a color factor),
 arising
from one-gluon-exchange diagram.
The 1-loop corrections (with massless and/or massive internal quarks) were computed in
\cite{Appelquist:es,Fischler:1977yf}.
The 2-loop correction (with massless internal quarks) was computed in
\cite{Peter:1996ig}.
The 2-loop correction due to massive internal quarks was computed
in \cite{Melles:2000dq}
(partly corrected in \cite{Recksiegel:2001xq}).\footnote{
Closely related is the computation of the 2-loop correction to the octet potential
 \cite{Kniehl:2004rk}.}
The logarithmic correction at 3-loop
originating from the ultrasoft scale
was first pointed out in \cite{Appelquist:es} and
computed in \cite{Brambilla:1999qa}.
Renormalization-group (RG) improvement of $V_{\rm QCD}(r)$ at
next-to-next-to-leading logarithmic order
was performed in \cite{Pineda:2000gz}.
A logarithmic contribution at ${\cal O}(\alpha_S^5)$
was computed in \cite{Brambilla:2006wp}.
The contributions of the massless quark loops to the 3-loop correction
were computed in \cite{Smirnov:2008pn}.
The only remaining correction at 3-loop order is the purely gluonic
contribution, which we compute in this paper.

For a long time, the perturbative QCD predictions of $V_{\rm QCD}(r)$
were {\it not} successful
in the intermediate distance region, relevant to the bottomonium and charmonium
states.
In fact, the perturbative series turned out to be poorly convergent at
$r \simgt 0.1~{\rm fm}$;
uncertainty of the series is so large that one could hardly obtain
meaningful prediction in this distance region.
Even if one tries to improve the perturbation series by
certain resummation prescriptions (such as RG improvement),
scheme dependence of the results turns out to be very large;
hence, one can neither
obtain accurate prediction of the potential in this region.
It was later pointed out that the large uncertainty of the perturbative
prediction can be understood as caused by
the ${\cal O}(\LQ)$ infrared (IR) renormalon
contained in $V_{\rm QCD}(r)$ \cite{Aglietti:1995tg}.

The situation has changed dramatically
since the discovery of the
cancellation of ${\cal O}(\Lambda_{\rm QCD})$
renormalons in the total energy of a static quark-antiquark pair
$E_{\rm tot}(r) \equiv V_{\rm QCD}(r) + 2m_{\rm pole}$ \cite{Pineda:id}.
Convergence of the perturbative series for $E_{\rm tot}(r)$
improved drastically and
much more accurate perturbative predictions
for the potential shape became available.
It was understood that a large uncertainty originating from
the ${\cal O}(\Lambda_{\rm QCD})$
renormalon in $V_{\rm QCD}(r)$ can be absorbed into
twice of the quark pole mass $2m_{\rm pole}$.
Once this is achieved, perturbative
uncertainty of $E_{\rm tot}(r)$ is
estimated to be much smaller.

Then it was readily recognized that perturbative convergence of
$V_{\rm QCD}(r)$ can be improved by adding a ($r$-independent) constant
at each order of the perturbative
expansion, since the ${\cal O}(\Lambda_{\rm QCD})$
renormalon is $r$-independent.
The conventional prescription to fix $V_{\rm QCD}(r)\to 0$ at $r\to\infty$
is not optimal as the convergence of perturbative series is worse
at larger $r$;
rather it is better to fix $V_{\rm QCD}(r)$ at some small distance.
As it turned out, $V_{\rm QCD}(r)$ becomes steeper at $r \simgt 0.1~{\rm fm}$
as the order of the
expansion is raised, hence convergence of perturbative series becomes
worse if we fix $V_{\rm QCD}(r)$ at $r\to\infty$.
This feature, that the perturbative potential becomes steeper than
the Coulomb potential as $r$ increases,
is understood, within perturbative QCD,
as an effect of the {\it running} of the strong coupling constant
\cite{Sumino:2001eh}.
In fact, several studies have
shown that perturbative
predictions for  $V_{\rm QCD}(r)$ agree well
with phonomenological potentials
and lattice calculations of $V_{\rm QCD}(r)$
in the intermediate distance region
\cite{Sumino:2001eh,Necco:2001xg,Recksiegel:2001xq,Pineda:2002se}.

The improvement of the situation opened up vast applications of
the QCD potential in heavy quarkonium physics
\cite{Brambilla:2004wf}.
For instance, higher--order computations of $V_{\rm QCD}(r)$
play crucial roles in precise determinations of
$m_c, m_b, m_t$ from the masses of charmonium, bottomonium
and (would-be) toponium states.
The 3--loop correction to $V_{\rm QCD}(r)$ is one of the missing parts
in these computations and also in
recent efforts to complete
next-to-next-to-next-to-leading order
corrections to heavy quark production
near threshold at $e^+ e^-$ colliders
\cite{Beneke:2005hg}.
Another application is a precise determination of $\alpha_S$,
from comparison of the perturbative prediction and lattice computations
of $V_{\rm QCD}(r)$ \cite{Sumino:2005cq}.

The static QCD potential is defined from an expectation value
of the Wilson loop as
\bea
&&
V_{\rm QCD}(r)
=
- \! \lim_{T \to \infty} \frac{1}{iT} \,
\ln \frac{\bra{0} {\rm{Tr\, P}}
e^{ i g \oint_{C} dx^\mu A_\mu }
\ket{0}}
{\bra{0} {\rm Tr} \, {\bf 1} \ket{0}}
\nonumber \\
&&~~~
=
\biggl(\frac{\mu^2 e^{\gamma_E}}{4\pi}\biggr)^{\epsilon}
\int \frac{d^d\vec{q}}{(2\pi)^d} \, e^{i \vec{q} \cdot \vec{r}}
\, \biggl[
-4 \pi  C_F \, \frac{\alpha_V(q)}{q^{2}}
\biggr]
\label{defalphaV}
\eea
where $q=|\vec{q}|$;
${C}$ is a rectangular loop of spatial extent $r$ and
time extent $T$.
The second equality defines the $V$-scheme coupling contant,
$\alpha_V(q)$, in momentum space.
We employ dimensional regularization with one temporal dimension
and $d = D-1 = 3-2\epsilon$ spatial dimensions.
A prefactor is included such that $\alpha_V(q)$ is
defined to be dimensionless;
$\gamma_E = 0.5772...$ denotes the Euler constant.

In perturbative QCD, $\alpha_V(q)$ is calculable in series
expansion of the strong coupling constant.
We denote the perturbative evaluation of
$\alpha_V(q)$ as
\bea
&&
\alpha_V^{\rm PT}(q)
= \alpha_S(\mu) \, \sum_{n=0}^{\infty} P_n(\ell ) \,
\biggl( \frac{\alpha_S(\mu)}{4\pi} \biggr)^n
\nonumber\\
\label{alfVPT}
\eea
with
\bea
&&~~~
\ell=\log (\mu/q).
\eea
Here, $\alpha_S(\mu)$ denotes the strong coupling constant
renormalized at the renormalization scale $\mu$,
defined in the modified minimal subtraction ($\overline{\rm MS}$) scheme;
$P_n(\ell)$ denotes an $n$-th-degree polynomial of $\ell$.
The RG equation of $\alpha_S(\mu)$ is
given by
\bea
\mu^2 \, \frac{d}{d\mu^2} \, \alpha_S(\mu) =
- \alpha_S(\mu) \sum_{n=-1}^{\infty} \beta_n
\biggl( \frac{\alpha_S(\mu)}{4\pi} \biggr)^{n+1},
\label{RGeq}
\eea
where $\beta_n$ represents the $(n+1)$-loop coefficient of the
beta function.\footnote{
In dimensional regularization and $\overline{\rm MS}$ scheme,
$\beta_{-1} \neq 0$ in the case $D\neq 4$.
}
For $n\leq 2$, the only part of the polynomial $P_n(\ell )$
that is not determined by the RG equation is $a_n \equiv P_n(0)$.
For $n\geq 3$, $P_n(\ell)$ includes IR divergences in terms
of poles of $\epsilon$ and assoicated logarithms, whose coefficients are
not determined by $\beta_n$.
At 3--loop order, we have
\bea
&&
{P}_3(\ell) =
a_3  + ( 6a_2\beta_0+
   4 a_1 \beta_1 +
   2 a_0 \beta_2) \ell
\nonumber\\ &&
~~~~~~~~
+
   (12 a_1 {\beta_0}^2
+ 10  a_0 \beta_0 \beta_1) \ell^2
+ 8 a_0 {\beta_0}^3  \ell^3 ,
\\ &&
a_3=\bar{a}_3+\frac{8}{3}\pi^2C_A^3\biggl(
\frac{1}{\epsilon}+6\,\ell
\biggr) .
\label{a3}
\eea
$C_F=(N_c^2-1)/(2N_c)$ and $C_A=N_c$ denote the eigenvalues of the quadratic
Casimir operators for the fundamental and adjoint representations, respectively,
of the color $SU(N_c)$ gauge group; $N_c=3$ in QCD.

The IR divergence is an artifact of the strict perturbative expansion
of $V_{\rm QCD}(r)$  in $\alpha_S$;
beyond naive perturbation theory, this IR divergence is absent and
regularized by the energy difference between color--singlet
and octet intermediate states.
The difference between $V_{\rm QCD}(r)$ and its perturbative expansion
$[ V_{\rm QCD}(r) ]_{\rm PT}$ can be treated systematically
within the effective field theory
``potential non--relativistic
QCD" \cite{Brambilla:2004jw}.
($[ V_{\rm QCD}(r) ]_{\rm PT}$ is obtained from $V_{\rm QCD}(r)$ if we replace
 $\alpha_V(q)$ in eq.~(\ref{defalphaV})
by $\alpha_V^{\rm PT}(q)$.)
This difference
\bea
\Bigl[ V_{\rm QCD}(r) \Bigr]_{\rm US} =V_{\rm QCD}(r)-\Bigl[ V_{\rm QCD}(r) \Bigr]_{\rm PT}
\eea
is given by contributions of ultra--soft (US) degrees
of freedom.\footnote{
$[ V_{\rm QCD}(r) ]_{\rm US}$ coincides with
$\delta E_{\rm US}(r)$ of \cite{Brambilla:2004jw} defined at the bare level.
This is because, if we perform strict expansion of $\delta E_{\rm US}(r)$ in $\alpha_S$,
it is expressed by scaleless integrals and therefore it vanishes.
}
In the region $r\ll \LQ^{-1}$, the leading--order contribution to
$[ V_{\rm QCD}(r) ]_{\rm US}$ in double expansion in $\alpha_S$ and $\log(\alpha_S)$ is
readily obtained from the result of \cite{Brambilla:1999qa} as
\bea
&&
\Bigl[ V_{\rm QCD}(r) \Bigr]_{\rm US, LO}=
\frac{C_FC_A^3\alpha_S^4}{24\,\pi \,r}\times
\biggl[ \,
\frac{1}{\epsilon} + 8 \log (\mu\, r)
\nonumber\\&&
~~~~~~~~~~~~~~~~~~~~~~~
-
2 \log \bigl({C_A \alpha_S} \bigr) +
\frac{5}{3} + 6\gamma_E
\biggr] .
\label{deltaEUS-pert}
\eea
\nonumber\\
Upon Fourier transform, $1/\epsilon$ and $\log \mu$ terms
of eqs.~(\ref{a3}) and (\ref{deltaEUS-pert})
cancel each other.
In general, at $r< \LQ^{-1}$, one may perform operator--product--expansion of
$[ V_{\rm QCD}(r) ]_{\rm US}$
as multi--pole expansion in $r$.
In this case,
non--perturbative contributions to $[ V_{\rm QCD} ]_{\rm US}$ are parametrized
in the form of non--local gluon condensates.

We may classify $\bar{a}_3$ in (\ref{a3}) according to the powers of the number of
flavors $n_l$ of the internal quarks:
\bea
\bar{a}_3 = n_l^3 \, \bar{a}_3^{(3)} + n_l^2 \, \bar{a}_3^{(2)} +
n_l \, \bar{a}_3^{(1)} + \bar{a}_3^{(0)} .
\eea
The purpose of this paper is to compute $\bar{a}_3^{(0)}$.

Let us describe our calculational procedure.
At tree--level and at 1--loop order, computation of $[ V_{\rm QCD}(r) ]_{\rm PT}$
is more or less trivial.
The 2--loop correction to $[ V_{\rm QCD}(r) ]_{\rm PT}$
is expressed in terms of 5 master integrals,
all of which are expressed in terms of $\Gamma$ function and
rational functions of $\epsilon$ \cite{Schroder:1999sg}.
Hence, we may easily obtain expansion coefficients in $\epsilon$
necessary for the 3--loop computation.

We first generate 3-loop Feynman diagrams for the scattering of static
quark and antiquark
using GRACE \cite{Ishikawa:1993qr} and QGRAF \cite{Nogueira:1991ex}.
There are about 20,000 diagrams;  we confirmed that the diagrams generated by
the two programs coincide.
Next step is to eliminate iterations of the lower-order potential at the
diagram level, which includes appropriate rearrangements of color factors
associated with diagrams;
we use the general algorithm developed in \cite{Gatheral:1983cz}.
This procedure eliminates diagrams which contain pinch singularities.
Subsequently the color factor  for each diagram is
simplified using the program \texttt{color} provided in
\cite{vanRitbergen:1998pn}.

Our computation is carried out in
Feynman gauge.
The loop integrals are classified according to different numerators and denominators.
At an early stage of the computation, we identify
those integrals which are trivially zero in dimensional regularization and
eliminate them.
To reduce the labor of computation, we collect
integrands with a common denominator
and cancel numerators against denominators as much as possible, by appropriately
expressing numerators in combinations of factors in the denominator.
After these processes, we were able to express the 3--loop correction to
$[ V_{\rm QCD}(r) ]_{\rm PT}$ in terms of about 1700
integrals.

Following the standard procedure of contemporary loop computations,
these integrals are expressed in terms of a small set of integrals (master integrals)
through the reduction procedure using integration-by-parts (IBP) identities \cite{Chetyrkin:1981qh}.
To carry out the reduction efficiently, we use the Laporta algorithm \cite{Laporta:2001dd}.
In addition to known techniques, we implement some improvement to
this reduction algorithm.
For instance, we temporarily assign a numerical value to $D$ and reduce integrals
to simpler ones using IBP identities.
Reduction process completes swiftly since manipulation of numerics is
considerably faster than symbolic
manipulation involving rational functions of $D$.
We retrace the reduction process and identify
a minimal set of necessary IBP identities for this reduction.
Then we reprocess the reduction (without assigning a numerical value to $D$)
using the minimal set of identities, after rearranging the order of these
identities optimally.
In the end, the 3--loop correction to
$[ V_{\rm QCD}(r) ]_{\rm PT}$ is expressed in terms of 40 master integrals.
All the processes are automatized and the integrals are
reduced one after another.
The reduction processes required roughly 3 weeks' CPU time of a comtemporary desktop
computer with 5~GB memory.

Out of 40 master integrals, 17 integrals can be expressed in terms of
$\Gamma$ function and rational functions of $D$.
The rest of the master integrals are expanded in Laurant series in $\epsilon$
and their expansion coefficients are evaluated analytically if possible and
 numerically otherwise.
(For some expansion coefficients, analytical values are available in the
literature.)
Numerical evaluation of the expansion coefficients are carried out in two ways:
(a) to evaluate Feynman parameter integrals using sector decomposition, and
(b) to evaluate integrals in Mellin-Barnes representation.
Typical relative
accuracy in numerical evaluation of
the expansion coefficients is of order $10^{-5}$.
Details of our computation will be described elsewhere.

Our final result reads
\bea
 \bar{a}_3^{(0)} = (502.22(12))\,C_A^3 + (-136.8(14))\, \frac{d_F^{abcd}d_A^{abcd}}{N_A}
 \label{ourresults}
\eea
with the color factor ${d_F^{abcd}d_A^{abcd}}/{N_A}=N_c(N_c^2+6)/48$
\cite{vanRitbergen:1998pn}.
For completeness, we combine
our result with that of \cite{Smirnov:2008pn}
and list the numerical values of $\bar{a}_3$, defined in eq.~(\ref{a3}),
for $N_c=3$ and $n_l=3,4,5$ in Tab.~\ref{table:a3bar}.
\begin{table}[b]
\begin{center}
\begin{tabular}{c|ccc}
\hline
\makebox[10mm]{$n_l$}&\makebox[20mm]{3}&\makebox[20mm]{4}&\makebox[20mm]{5} \\
\hline
$\bar{a}_3$ & 5199(3) & 3160(3)  & 1460(3)      \\
\hline
\end{tabular}
\caption{\small
Numerical values of $\bar{a}_3$, defined in eq.~(\ref{a3}),
for different values of $n_l$.
}
\label{table:a3bar}
\end{center}
\end{table}

At every stage of the computation we performed numerous cross checks.
At every step we have written (at least) two independent programs and
checked that results mutually agree.
We derived many relations among different types of integrals and checked that,
when the integrals are expressed by master integrals, these relations are
satisfied.
Renormalizability of the QCD potential
with the known renormalization constant of the strong coupling constant, as well as
reproduction of known IR divergence, provide non--trivial cross checks.
We have reproduced $\bar{a}_3^{(3)}$, $\bar{a}_3^{(2)}$, and
the coefficient of $C_F^2$ in $\bar{a}_3^{(1)}$ \cite{Smirnov:2008pn}
analytically.
We also computed the coefficients of $C_A^2$, $C_AC_F$ and $d_F^{abcd}d_F^{abcd}/N_A$
in $\bar{a}_3^{(1)}$
numerically and confirmed that our results agree with those of \cite{Smirnov:2008pn} within the estimated errors.
The last comparison provides a strong cross check on correctness and
accuracy of our result (\ref{ourresults}), since the expansion
coefficients necessary to compute the result
(\ref{ourresults}) are common to the ones necessary to compute
$\bar{a}_3^{(3)}$, $\bar{a}_3^{(2)}$, $\bar{a}_3^{(1)}$ except for a single
coefficient.

Now we compute $V_{\rm QCD}(r)$,
as given by
the sum of $[V_{\rm QCD}(r)]_{\rm PT}$ and
$[ V_{\rm QCD}(r) ]_{\rm US}$,
including all the corrections up to ${\cal O}(\alpha_s^4)$ and
${\cal O}(\alpha_s^4\log\alpha_S)$.
Namely we use the series expansion (\ref{alfVPT})
up to $n=3$ for the former and
eq.~(\ref{deltaEUS-pert}) for the latter.

\begin{figure}[t]
\hspace*{-30mm}
\includegraphics[width=9cm]{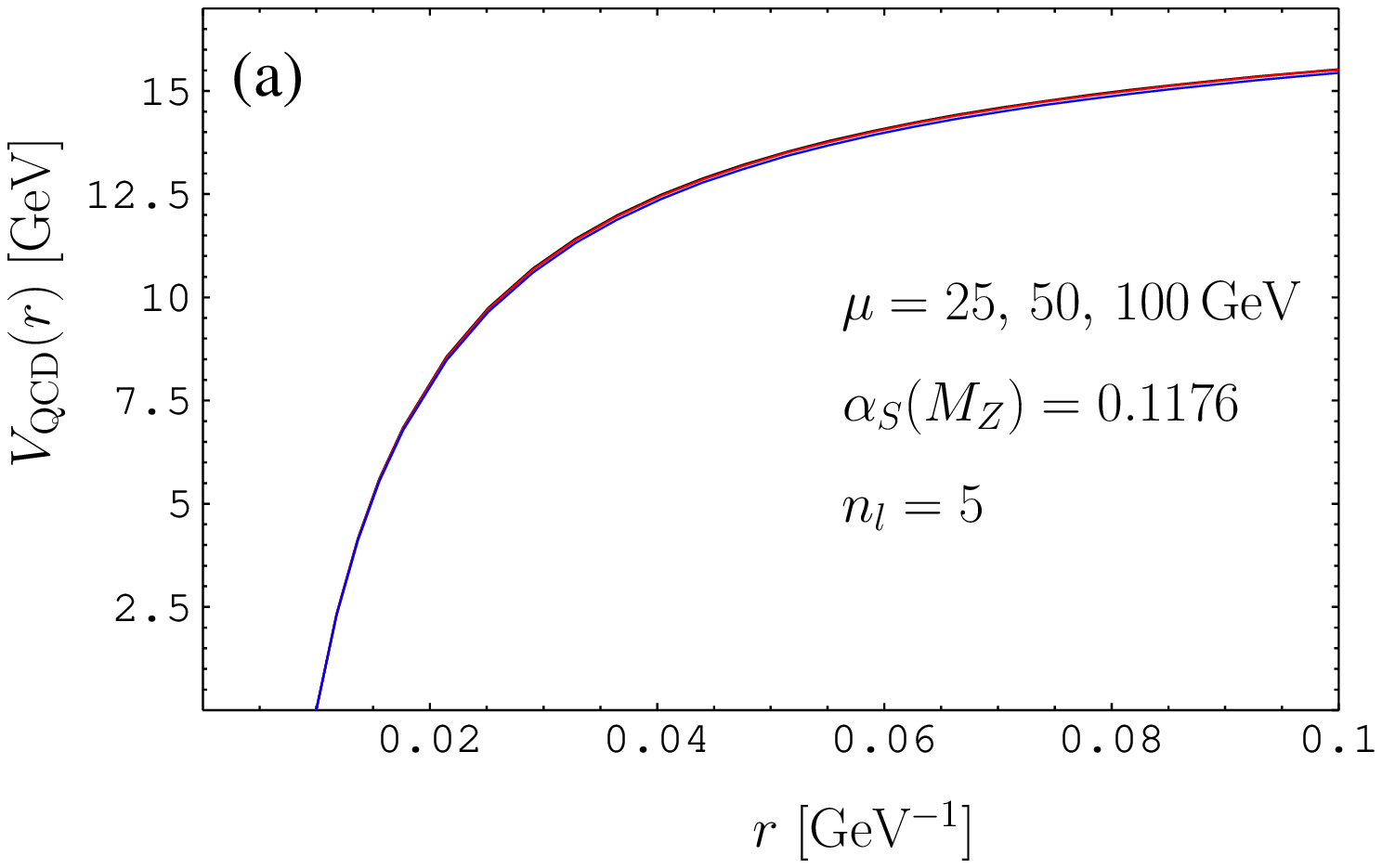}
\vspace*{1mm}
\\
\hspace*{-20mm}
\includegraphics[width=9cm]{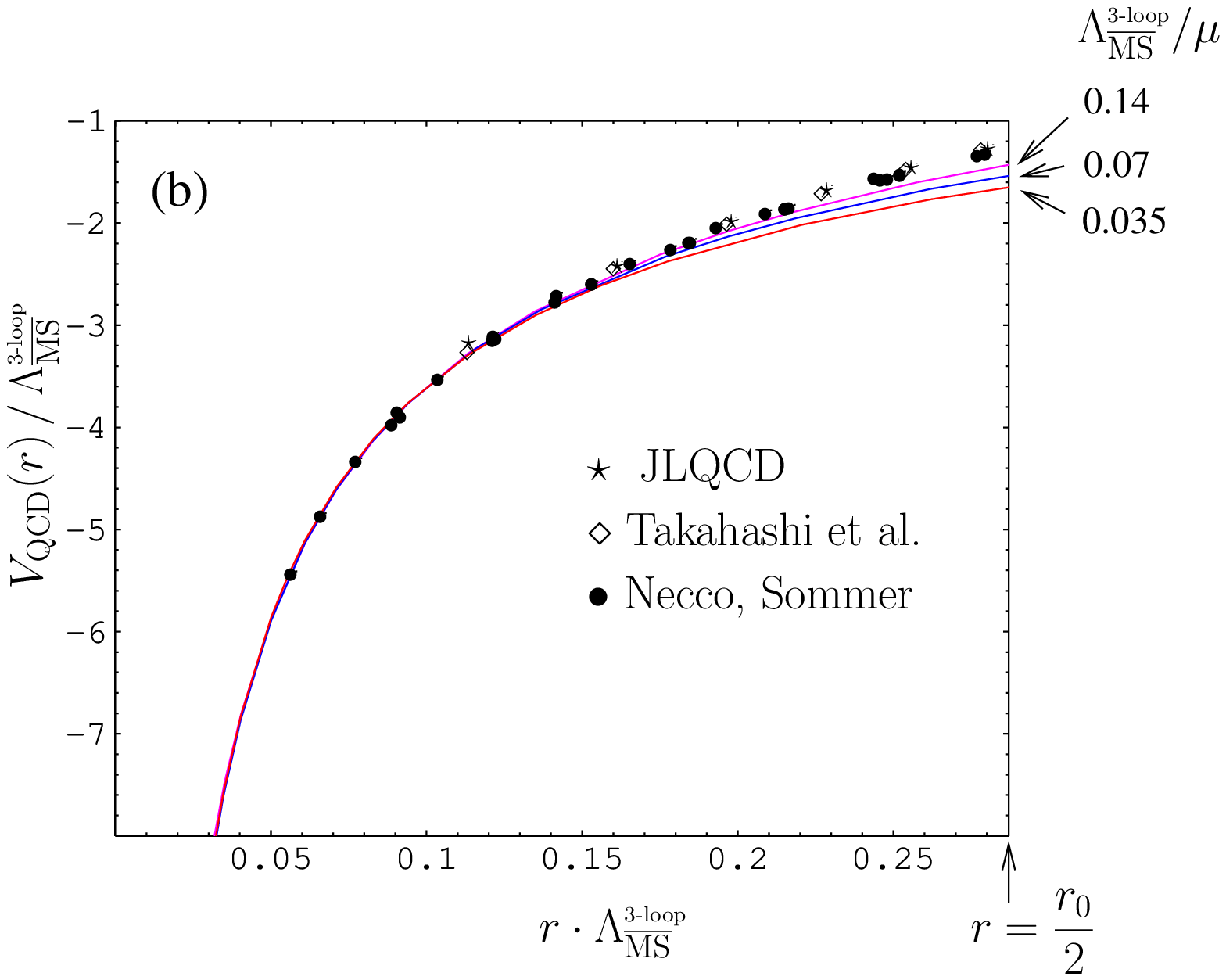}
\caption{\small
$V_{\rm QCD}(r)=[V_{\rm QCD}(r)]_{\rm PT}+[ V_{\rm QCD}(r) ]_{\rm US}$
up to ${\cal O}(\alpha_s^4)$ and
${\cal O}(\alpha_s^4\log\alpha_S)$:
(a) In the toponium region
3 lines, corresponding to
$\mu=25$, 50 and 100~GeV,  are plotted with $n_l=5$;
(b) Comparison with the lattice computations in the quenched
approximation \cite{Takahashi:2002bw,Necco:2001xg}.
We set $n_l=0$. The distance region
corresponds roughly to the size of $\Upsilon(1S)$ state.
\label{Fig}
}
\end{figure}

In Fig.~\ref{Fig}(a) we plot our prediction for
$V_{\rm QCD}(r)$ in the distance region
corresponding to (would-be)
toponium states.
3 lines are plotted, corresponding to
$\mu=25$, 50 and 100~GeV, with $n_l=5$ and $\alpha_S(M_Z)=0.1176$.
We added a constant to each prediction such that it takes a common value at
$r=0.01~{\rm GeV}^{-1}$.
The difference of the 3 lines are hardly visible, showing
stability of the prediction.

In Fig.~\ref{Fig}(b) we compare our prediction with the lattice
data in the quenched approximation \cite{Takahashi:2002bw,Necco:2001xg}.
Accordingly we set $n_l=0$.
We used the central value of
$r_0 \Lambda_{\overline{\rm MS}}^{\mbox{\scriptsize 3-loop}}
=0.574\pm 0.042
$ \cite{Sumino:2005cq}
to fix the relation between the lattice scale
 and
$\Lambda_{\overline{\rm MS}}^{\mbox{\scriptsize 3-loop}}$,
where $r_0$ denotes the Sommer scale.
Hence, the only adjustable parameters in our comparison are
$r$--independent constants to be added to the potentials,
whose values are chosen such that
all the potentials coincide at
$r \Lambda_{\overline{\rm MS}}^{\mbox{\scriptsize 3-loop}} = 0.1$.
It is customary to interpret $r_0=0.5$~fm
when comparing this scale to one of the real world.
Roughly the potential shape in the displayed
range $r<r_0/2$ accounts for
formation of the $\Upsilon(1S)$ state.
We plot 3 lines with the scale choices
$\Lambda_{\overline{\rm MS}}^{\mbox{\scriptsize 3-loop}}/\mu
= 0.14$, 0.07 and 0.035.
(The corresponding values of $\alpha_S(\mu)$ are
0.216, 0.165 and 0.135, respectively.)
There is a small but visible dependence on the scale.
The level of agreement with the lattice data shows that
our prediction of the potential at this order is good enough to warrant
quantitative description of the nature of the $\Upsilon(1S)$ state.
We confirm the observation that, as we include higher--order corrections,
agreement of the perturbative prediction and lattice computations
improves up to larger distances.
According to the analyses in
\cite{Sumino:2001eh,Sumino:2005cq}, we anticipate that the agreement would
get even better if we resum logarithms via RG, or, appropriately
choose the scale $\mu$ as a function of $r$, {provided}
that the IR renormalon is subtracted.

\medbreak
The work of Y.S.\ is supported in part by Grant-in-Aid for
scientific research No.\ 17540228 from
MEXT, Japan.


\end{document}